\documentclass{sf2a-conf2011}
\usepackage{graphicx}
\usepackage{hyperref}
\usepackage[]{natbib}

\begin{document}
\TitreGlobal{SF2A 2011}
%
\title{Calibration of the Gaia RVS from ground-based observations of candidate standard stars}
\author{L. Chemin}\address{Laboratoire d'Astrophysique de Bordeaux, UMR 5804 (CNRS, Universit\'e Bordeaux 1), 33271 Floirac Cedex, France}
\author{C. Soubiran$^1$}
\author{F. Crifo}\address{Observatoire de Paris, GEPI, UMR 8111 (CNRS , Universit\'e Denis Diderot Paris 7), 92195 Meudon, France}
\author{G. Jasniewicz}\address{GRAAL, UMR5024, (CNRS, Universit\'e Montpellier 2), 34095 Montpellier Cedex 05, France}
\author{D. Katz$^2$}
\author{D. Hestroffer}\address{Observatoire de Paris, IMCCE, UMR8028 (CNRS, Universit\'e Pierre \& Marie Curie Paris 6),  75014, Paris, France}
\author{S. Udry}\address{Observatoire de Gen\`eve, 51 Ch. des Maillettes, 1290 Sauverny, Switzerland}
\runningtitle{Calibration of the Gaia RVS from ground-based observations of candidate standard stars}
%
\setcounter{page}{237}
\index{L. Chemin}
\index{C. Soubiran}
\index{F. Crifo}
\index{G. Jasniewicz}
\index{D. Katz} 
\index{D. Hestroffer}
\index{S. Udry}

\maketitle
\begin{abstract}
The Radial Velocity Spectrometer (RVS) on board of Gaia will perform a large 
spectroscopic survey to determine the radial velocities of some $1.5 \times 10^8$ stars. 
We present the status of ground-based observations of a sample of 1420 candidate 
standard stars designed to calibrate the RVS. 
 Each candidate star has to be observed several times before Gaia launch 
 (and at least once during the mission) to ensure that its radial velocity  
 remains stable during the whole mission. Observations are performed with the high-resolution spectrographs SOPHIE, 
 NARVAL and CORALIE, completed with archival data of the ELODIE and HARPS instruments. The analysis shows that about 7\% 
 of the current catalogue exhibits variations larger than the adopted threshold of 
 300 m s$^{-1}$. Consequently, those stars 
 should be rejected as reference targets, due to the expected accuracy of the Gaia RVS. Emphasis is also put here on our  observations of bright 
 asteroids to calibrate  the ground-based velocities by a direct comparison with  
 celestial mechanics. It is shown that the radial velocity zero points of SOPHIE, 
 NARVAL and CORALIE are consistent with each other, within the uncertainties. 
 Despite some scatter, their temporal variations 
 remain small with respect to our adopted stability criterion.

\end{abstract}
\section{Generalities}

The RVS is a slitless spectrograph   whose spectral domain is 847-874 nm and resolving power $R \sim 11500$. 
The expected accuracy is 1 km s$^{-1}$ for F0 to K0 stars brighter than V$=13$, and for K1 to K4 stars brighter than V$=14$.

The main scientific objectives of RVS are the chemistry and dynamics of the Milky Way, the 
detection and characterisation of multiple systems and variable stars \citep[for more details, see][]{wil05}.
Those objectives will be achieved from a spectroscopic survey of:  

\begin{itemize}
\item    Radial velocities ($\sim 150 \times 10^6$ objects, V $\le 17$)
\item    Rotational velocities ($\sim 5 \times 10^6$ objects, V $\le 13$)
\item    Atmospheric parameters ($\sim 5 \times 10^6$ objects, V $\le 13$)
\item    Abundances ($\sim 2 \times 10^6$ objects, V $\le 12$)
\end{itemize}

Each star will be observed $\sim 40$ times on average by RVS over the 5 years of the mission.
 
\section{Calibration of the Gaia RVS}

Because the RVS has no calibration module on board, the zero point of its radial velocities has to be determined 
from reference sources. Ground-based observations of a large sample of well-known, 
stable reference stars as well as of asteroids are thus critical for  
the calibration of the RVS. A sample of 1420 candidate standard stars has been established \citep{cri09,cri10} and has to be validated 
by high spectral resolution observations. 

Two measurements per candidate star are being made before Gaia is launched (or one, depending on already available archived data).  
Another measurement will occur during the mission. The measurements will allow us to check the temporal stability of radial velocities, 
and to reject any targets with significant RV variation.

\section{Status of observations of stars}
 The ongoing observations are performed with three high spectral 
resolution spectrographs:
\begin{itemize}
\item SOPHIE on the 1.93-m telescope at Observatoire de Haute-Provence, 
\item NARVAL on the T\'elescope Bernard Lyot at Observatoire Pic-du-Midi,
\item CORALIE on the Euler swiss telescope at La Silla.
\end{itemize}

As of June 2011 we have observed 995 distinct candidates with SOPHIE, CORALIE and NARVAL. The detailed observations per instrument are:
\begin{itemize}
\item 691 stars (1165 velocities) with SOPHIE
\item 669 stars (945 velocities) with CORALIE 
\item  93 stars (98 velocities) with NARVAL
\end{itemize}

\begin{figure}[b!]
 \includegraphics[width=0.5\columnwidth]{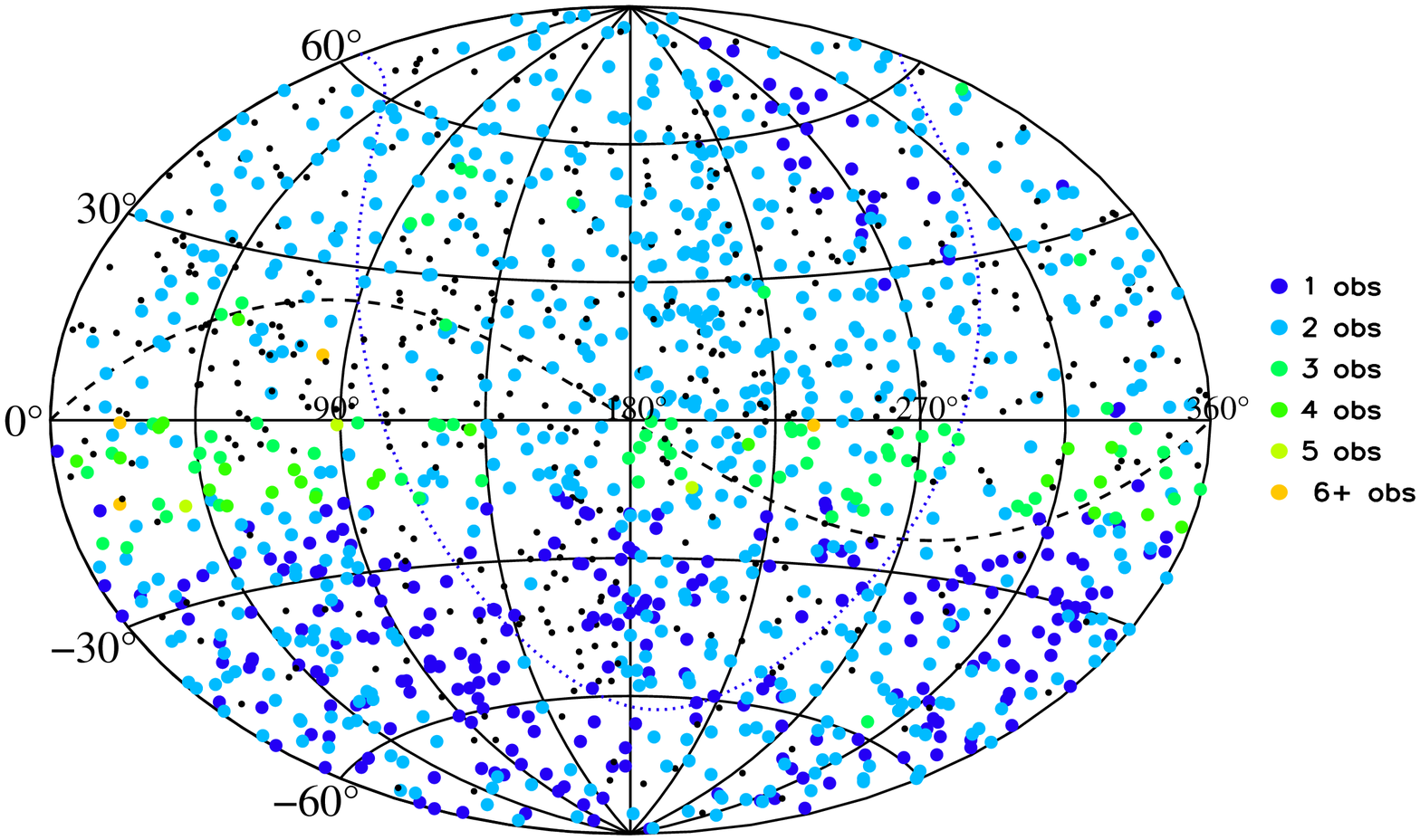}\includegraphics[width=0.5\columnwidth]{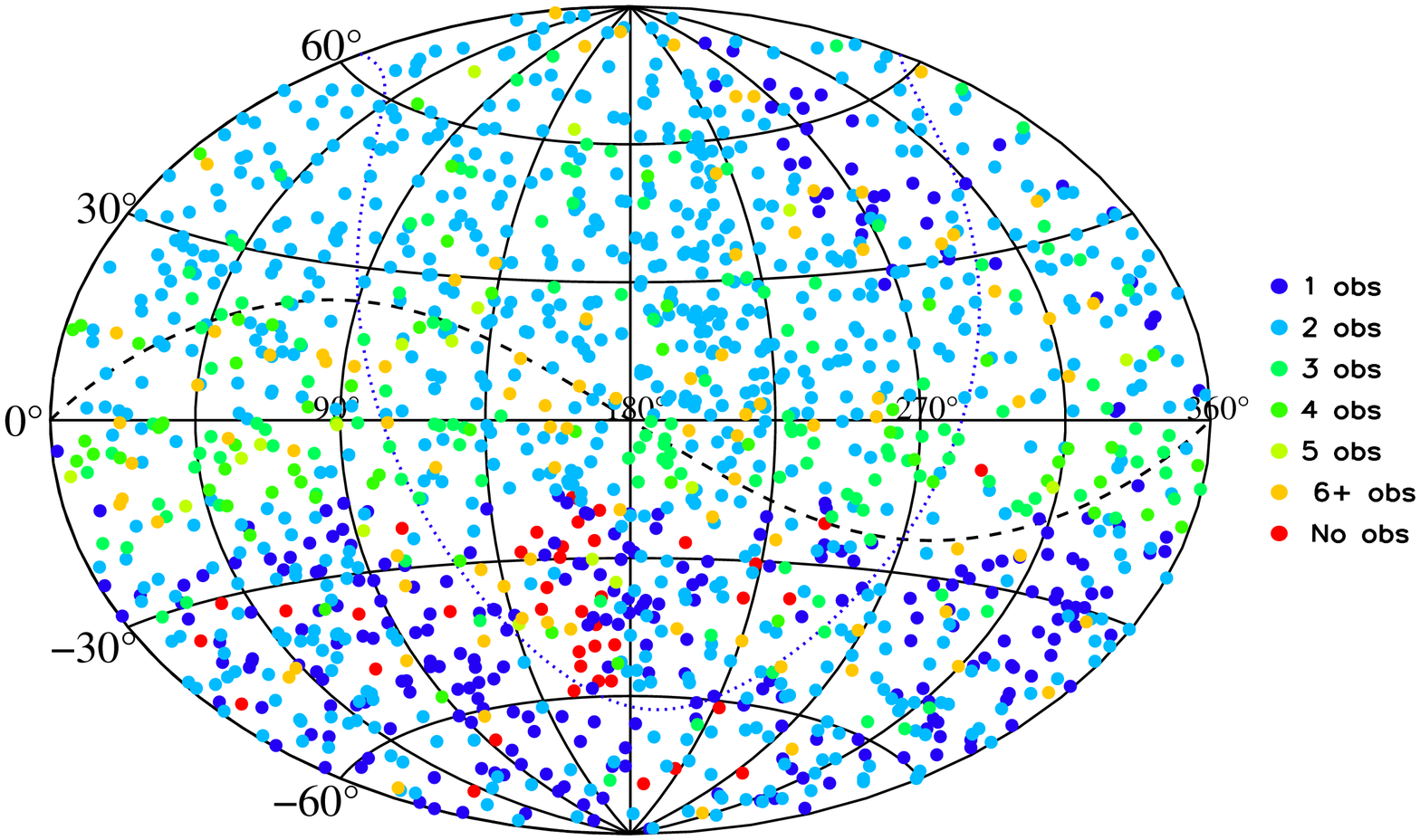}      
  \caption{\textbf{Left panel:} Current status of the number of observations available per candidate standard star  observed 
  with the SOPHIE, CORALIE and NARVAL instruments (995 distinct stars). 
  Black dots indicate the locations of the 425 remaining sources from the sample of 1420 candidates. 
    \textbf{Right panel:} Same as in left panel, but including archival data of the ELODIE and HARPS instruments as well. 
   The maps are represented in the equatorial frame. The Ecliptic plane is shown as a dashed line and the Galactic plane as a dotted line. }
  \label{chemin:fig1}
\end{figure}

Figure~\ref{chemin:fig1} (left-hand panel) represents the spatial distribution in the equatorial frame of the sample 
and the number of measurements per object we have done so far with the three instruments.  

In addition to those new observations, we use  radial velocity measurements available from the  spectroscopic archives of two other high-resolution instruments: ELODIE,  
which is a former OHP spectrograph, and HARPS which is currently observing at the ESO La Silla 3.6-m telescope. 
The archived data allow us to recover 1057 radial velocities for 292 stars (ELODIE) and 1289 velocities for 113 stars (HARPS).  

  Figure~\ref{chemin:fig1} (right-hand panel) summarizes the status of the total number of measurements for the  sample of 1420 candidate stars performed 
  with all five instruments.  

\section{How stable are the radial velocities of our candidates?}

\begin{figure}[h!]
 \begin{center}
 \includegraphics[width=10cm]{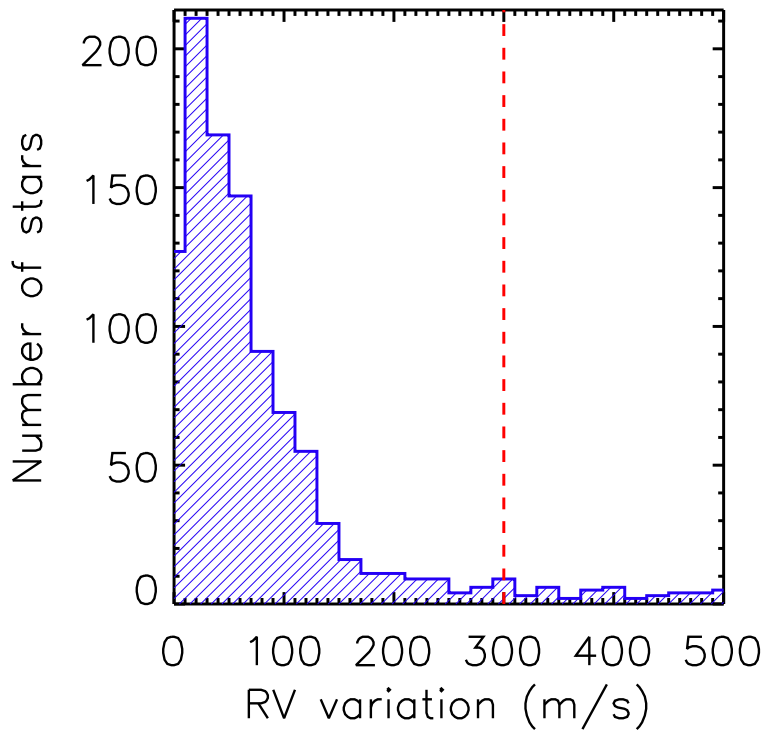} 
  \caption{Distribution of variations of radial velocities of candidate standard stars having at least two RV measurements separated by  100 days or more. 
  A dashed line shows the adopted 300 m s$^{-1}$ stability threshold.}
  \label{chemin:fig2}
  \end{center}
\end{figure}

We have derived the variation of radial velocity of each star for which we have at least two velocity measurements separated 
 by an elapsed time of at least 100 days.  These stars represent  a subsample of 1044  among 1420 targets. 
 The variation is defined as the difference between the maximum and minimum velocities, as reported in the frame of the SOPHIE spectrograph. 
  Its distribution is displayed in Figure~\ref{chemin:fig2}. 
  
  A candidate is considered as a reference star for the RVS calibration when its radial velocity does not vary by more than an adopted threshold of 300 m s$^{-1}$. 
  Such a threshold has been defined to satisfy the condition that the variation of the RV of a candidate must be well smaller than 
  the expected RVS accuracy  (1 km s$^{-1}$ at best for the brightest stars). As a result, we find $\sim 7\%$ 
  of the 1044 stars exhibiting a variation larger than 300 m s$^{-1}$, as derived from   available measurements performed to date. 
   Those variable stars will have to be rejected from the list of standard stars.
  Note   that about 75\% of the 1044 stars have very stable RV, at a level of variation smaller than 100 m s$^{-1}$.  

\section{Spectral observations of asteroids}

Observations of asteroids are very important for the radial velocity calibration. Indeed they will be used to determine the 
zero-points of the RVs measured with SOPHIE, CORALIE and NARVAL (as well as the Gaia-RVS zero-point). 
Those goals  will be achieved by comparing the spectroscopic RVs of asteroids from 
ground-based measurements with theoretical kinematical RVs from celestial mechanics.  
 The theoretical RVs are provided by IMCCE and are known with an accuracy better than 1 m s$^{-1}$. 
 About 280 measurements of 90 asteroids have been done so far. 
 
\begin{figure}[t!]
\centering
 \includegraphics[width=8cm]{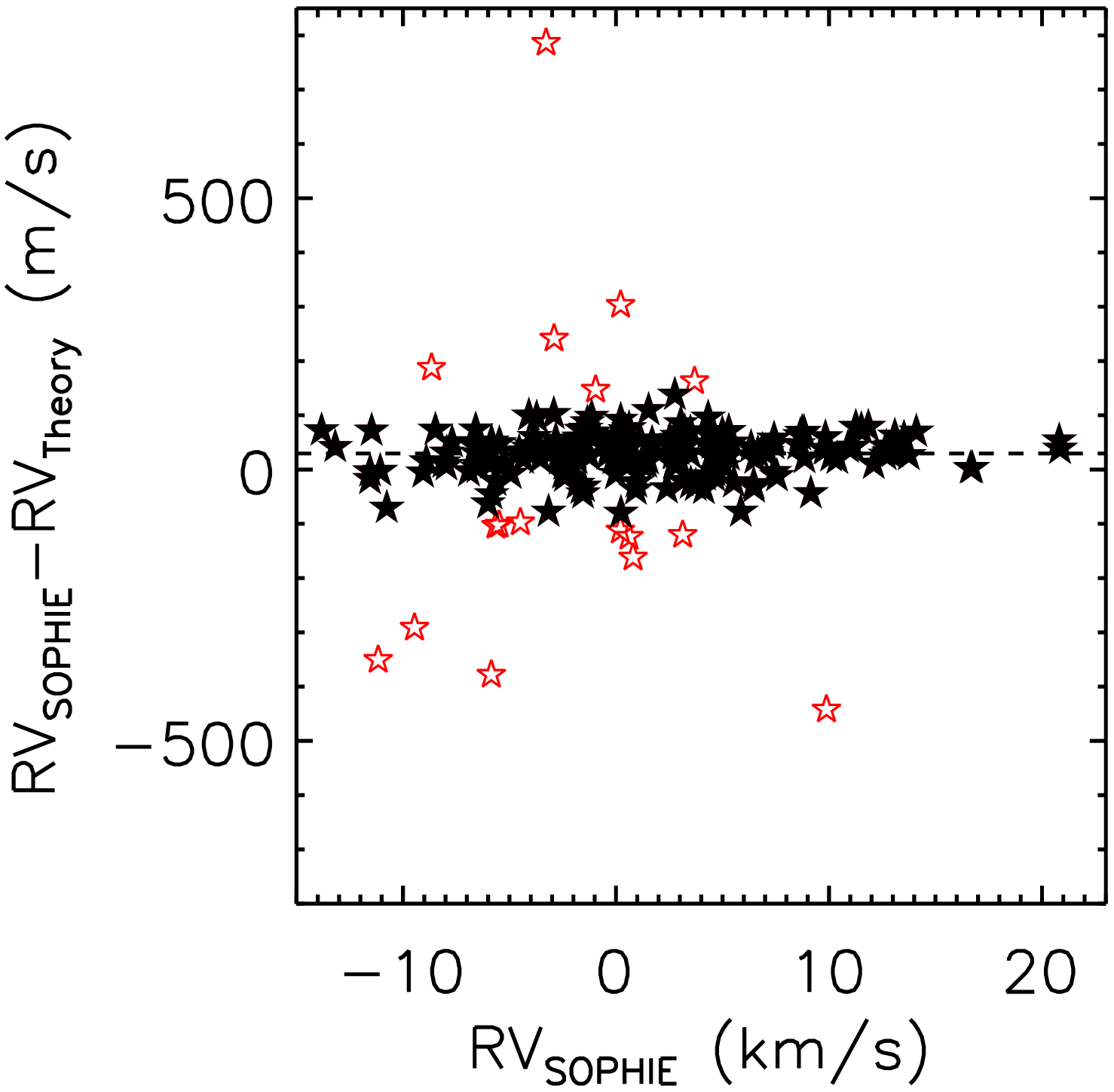}\includegraphics[width=8cm]{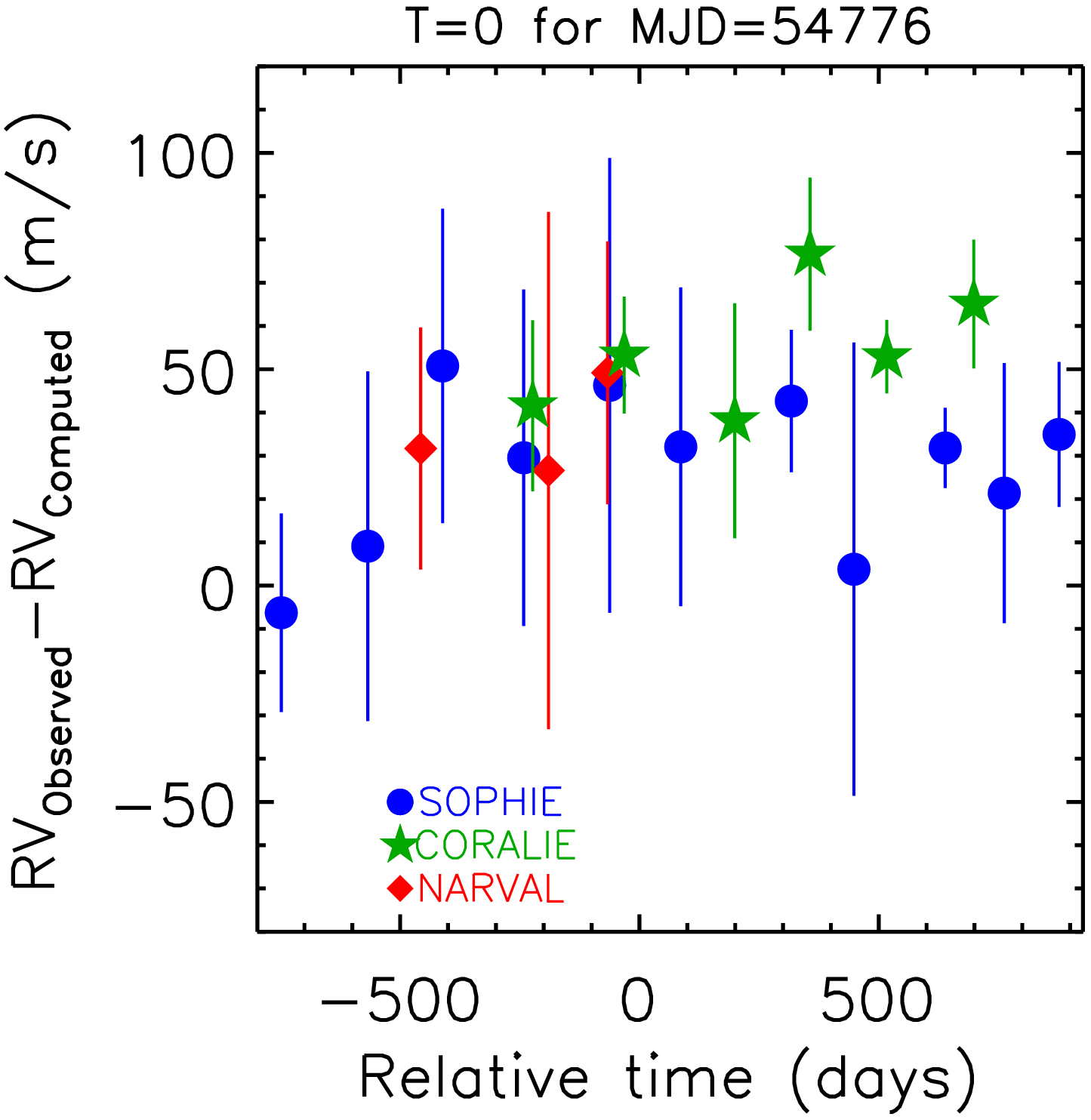}      
  \caption{Radial velocities of asteroids. \textbf{Left panel:} Residual velocities (observed minus computed) of asteroids as a function of their observed velocities 
  (SOPHIE observations only). Red symbols are points deviant by more than 3$\sigma$. A horizontal dashed line represents the mean residual velocity 
  of 30 m s$^{-1}$. \textbf{Right panel:} Comparison of residual velocities of asteroids for SOPHIE, CORALIE and NARVAL as a function of time for the various 
  observing runs.}
  \label{chemin:fig3}
\end{figure}

As an illustration, Figure~\ref{chemin:fig3} (left-hand panel) displays the residual velocity (observed minus computed RVs) of asteroids observed by the 
SOPHIE instrument as a function of the observed RVs.  The average residual of asteroids observed with SOPHIE is 30 m s$^{-1}$ and the scatter is 38 m s$^{-1}$.

In Figure~\ref{chemin:fig3} (right-hand panel) we also show the variation of the residual RVs with time. It nicely shows how stable the RVs are 
as a function of time. The residual RVs are relatively constant within the quoted errors. 
The error-bars represent the dispersion of all measurements performed at each observational run. Their amplitude is mainly related to the conditions of 
observations that differ from one session to another (in particular the moonlight contamination). Though significant (between 10 and 50 m s$^{-1}$) 
those error-bars remain smaller than our target stability criterion of 300 m s$^{-1}$, which will enable us to determine correctly the RV zero point of each instrument.

Note also we have   verified that intrinsic properties of asteroids (e.g. their size, shape, rotation velocity, albedo, etc...) 
have negligible systematic impacts on the determination of RVs zero points for the spectrographs. 
From now, observations of asteroids shall be performed with reduced moonlight contamination.

\begin{acknowledgements}
We are very grateful to the AS-Gaia, the PNPS and PNCG  for the financial support of the observing campaigns and the help in this project.
\end{acknowledgements}


\begin{thebibliography}{}

\bibitem[Crifo et al.(2009)]{cri09} Crifo F., Jasniewicz G., Soubiran C.,  et al., 2009, SF2A 2009 conference proceedings, 267
\bibitem[Crifo et al.(2010)]{cri10} Crifo F., Jasniewicz G., Soubiran C.,  et al., 2010, A\&A, 524, A10
\bibitem[Wilkinson et al.(2005)]{wil05} Wilkinson M.,  et al., 2005,  MNRAS, 359, 1306
 \end{thebibliography}
\end{document}